\documentclass[aps,prl,twocolumn,superscriptaddress]{revtex4}
\usepackage{amsmath}
\usepackage{epsfig}
\usepackage{rotating,dcolumn}
\begin{document}

\title{Magnetoelectric MnPS$_3$ as a new candidate  for  ferrotoroidicity}

\author{E. Ressouche}
\email[]{ressouche@ill.fr}
\affiliation{INAC/SPSMS-MDN, CEA/Grenoble, 17 rue des Martyrs, 38054 Grenoble Cedex 9, France}
\author{M. Loire} \affiliation{Institut N\'eel, CNRS \& Universit\'e Joseph Fourier, 
BP166, 38042 Grenoble, France}
\author{V. Simonet}
\affiliation{Institut N\'eel, CNRS \& Universit\'e Joseph Fourier,
BP166, 38042 Grenoble, France}
\author{R. Ballou} \affiliation{Institut N\'eel, CNRS \& Universit\'e Joseph Fourier, 
BP166, 38042 Grenoble Cedex 9, France}
\author{A. Stunault}
\affiliation{Institut Laue-Langevin, BP 156, 38042 Grenoble Cedex 9, France}
\author{A. Wildes}
\affiliation{Institut Laue-Langevin, BP 156, 38042 Grenoble Cedex 9, France}

\date{\today}

\begin{abstract}

We have revisited the magnetic structure of manganese phosphorus trisulfide MnPS$_3$ using neutron diffraction and polarimetry. MnPS$_3$ undergoes a transition towards a collinear antiferromagnetic order at 78 K. The resulting magnetic point group breaks both the time reversal and the space inversion thus allowing a linear magnetoelectric coupling. Neutron polarimetry was subsequently used to prove that this coupling provides a way to manipulate the antiferromagnetic domains simply by cooling the sample under crossed magnetic and electrical fields, in agreement with the non-diagonal form of the  magnetoelectric tensor. In addition, this tensor has in principle an antisymmetric part that results in a toroidic moment and provides with a pure ferrotoroidic compound. 

\end{abstract}


\maketitle 

Recent years have seen a renewed interest for magnetoelectricity  and a fast increasing number of studies devoted to multiferroics materials \cite{fiebig}. These compounds are defined as presenting at least two ferroic orders amongst the three widely known: ferroelasticity, ferroelectricity and ferromagnetism.  Although their antiferro- counterparts seem to be less interesting for possible technological applications, they are also often considered in the above classification. Recently, a fourth ferroic order, ferrotoroidicity, has been proposed as an order parameter associated with the breaking of both time reversal and space inversion whereas ferroelasticity is symmetric with respect to these two operations, ferromagnetism only breaks the time reversal and ferroelectricity the space inversion \cite{fiebig,dubovik,gorbatsevich1,sannikov,ederer,schmid}.

Ferrotoroidicity is intrinsically associated with the linear magnetoelectric (ME) coupling from which a magnetization (electric polarization) can be proportionally induced by an electric (magnetic) field \cite{landau}. The toroidic moment can be seen as generated by a vortex of magnetic moments and is defined as ${1\over 2}\sum_i \vec r_i\times \vec M_i$ (with $i$ labelling the magnetic atoms in the unit cell). It is proportional to the antisymmetric part of the ME tensor $\alpha^{ME}$, hence present only when this tensor has non-diagonal terms \cite{gorbatsevich2}. 

Multiferroics usually exhibit complex magnetic orders, often driven by magnetic frustration, with e.g. cycloids or incommensurate modulated phases.  However, for a linear ME coupling to be allowed in a centrosymmetric crystal, a necessary condition is a magnetic structure with a zero propagation vector which loses the inversion center below the ordering temperature \cite{landau}.  A number of  oxides have been identified to fulfill this condition and to be  potential magnetoelectrics \cite{Tables}. The most famous and studied example is Cr$_{2}$O$_{3}$, with a linear diagonal ME tensor associated with a collinear antiferromagnetic order without any spontaneous electric polarization \cite{djialoshinsky,astrov,rado}. The transition temperature is higher than room temperature which is promising for potential applications:  for instance, the switching of the exchange bias has been proven in a Cr$_{2}$O$_{3}$/(CoPt)$_3$ antiferromagnetic/ferromagnetic heterostructure. Performing ME annealing \cite{rado}, {\it i.e.} cooling through the N\'eel temperature in electric ($E$) and magnetic ($H$) fields (parallel or anti-parallel), the antiferromagnetic domains population was unbalanced in the magnetoelectric layer \cite{borisov1}. In Cr$_{2}$O$_{3}$, the antiferromagnetic domains are the two so-called 180$^{\circ}$  domains with moments in opposite directions (time reversal conjugates). They have different energies when submitted to an electric and magnetic field, the difference being proportional to the corresponding elements of the magnetoelectric tensor $\Delta G=2\alpha^{ME}_{ij}E_iH_j$. The work of Borisov {\it et al.} on exchange bias \cite{borisov1} underlines the importance of domains which form upon a phase transition to an (anti)ferroic phase, and of their manipulation. 

Up to now, two techniques have proven to be able to probe such 180$^{\circ}$ antiferromagnetic domains: second harmonic generation optical spectroscopy \cite{aken1} and neutron polarimetry \cite{brown1}. This last technique measures the domain population and has been used for instance in Cr$_{2}$O$_{3}$, MnGeO$_3$ and LiCoPO$_4$ to evidence the imbalance of domains through cooling in $E$ and $H$ \cite{brown1,brown2}. 

In the search for new magnetoelectric candidates, we have investigated MnPS$_3$ by neutron diffraction and neutron polarimetry. Its main structural and magnetic properties are known since the 1980s \cite{ouvrard,kurosawa,Okuda}, but this compound was never reported to show any magnetoelectric coupling. MnPS$_3$ (manganese phosphorus trisulfide, sometimes referred to as manganese thiophosphate) belongs to a family of compounds mainly studied for their intercalation properties \cite{brec}. It is a lamellar compound, crystallizing in the monoclinic space group C2/m. MnPS$_3$ is a highly resistive broad band semiconductor with a gap close to 3 eV and is optically transparent with a green colour \cite{grasso}. The transition metal ions Mn$^{2+}$ with spin 5/2, responsible for the magnetic properties, form a honeycomb lattice in the ($a$,$b$) plane. The weak interlayer coupling mediated by the S atoms, was proposed to be purely of Van der Waals origin. However, the analysis of the transition occurring at 78 K towards an antiferromagnetic collinear phase, suggests an interplane exchange associated with some degree of metal-ligand covalency \cite{wildes1,wildes2}. 

In this article, we report on the evidence and characterization by neutron polarimetry of linear ME coupling in MnPS$_3$ and of its influence on the antiferromagnetic domain populations. We discuss these results, in particular the consequence of the symmetry of the ME tensor with respect to ferrotoroidicity.

 A MnPS$_3$ crystal, grown by the method reported in reference \cite{wildes1}, has been studied by neutron diffraction at the Institut Laue-Langevin. The crystal and magnetic structures were first checked on the four-circle CEA-CRG diffractometer D15 using a wavelength of 1.174 \AA, at low temperature, in a closed-cycle refrigerator. The crystal structure was confirmed to be C2/m with unique $b$ axis with cell parameters a=6.05(1) \AA, b=10.52(3) \AA, c=6.80(2) \AA, $\alpha$=$\gamma$=90 $^{\circ}$ and $\beta$=107.3(2) $^{\circ}$ at 90 K. The extinction  turned out to be completely negligible. 

Spherical neutron polarimetry experiments were performed on the hot neutron beam diffractometer D3 with a wavelength of 0.84 \AA, using CRYOPAD. The technique consists in setting the polarization of the incident beam along different directions, and to analyze the polarization vector of the scattered beam. More detailed descriptions  of the technique can be found in ref. \cite{brown1,brown2}. Such measurements usually yields for each Bragg reflection a 3x3 matrix $P_{ij}$, with $i,j=X, Y, Z$, that relates the scattered polarization to the incoming one.  For each reflection, $X$ is parallel to the scattering vector, $Z$ is vertical and $Y$ completes the right handed set. The $P_{ij}$ matrix elements contain all the information on the arrangement of the magnetic moments and on the different magnetic domains if present. For this experiment the crystal was mounted with the $a$ axis vertical (defining $Z$), to perform measurements in the ($b^*$, $c^*$) scattering plane. 

In a first step, the magnetic structure was refined from the integrated intensities collected on D15 at 10 K. The resulting magnetic structure was found to agree  with the published one \cite{kurosawa}. Below T$_N$, the antiferromagnetic phase is characterised by a zero propagation vector. The four magnetic atoms (0, 0.3327, 0), (0, 0.6673, 0), (0.5, 0.8327, 0) and (0.5, 0.1673, 0) on the 4$g$ Wyckoff site have their magnetic moment coupled following a $+-+-$ sequence, {\it i.e.} each Mn$^{2+}$ is antiferromagnetically coupled with its nearest neighbors in the ($a$, $b$) plane, and the coupling between adjacent planes is ferromagnetic. The magnetic moments were reported to lie along $c^*$ \cite{kurosawa}.  

\begin{figure}[h]
\begin{minipage}[h]{.47\linewidth}
\includegraphics[width=4.0cm]{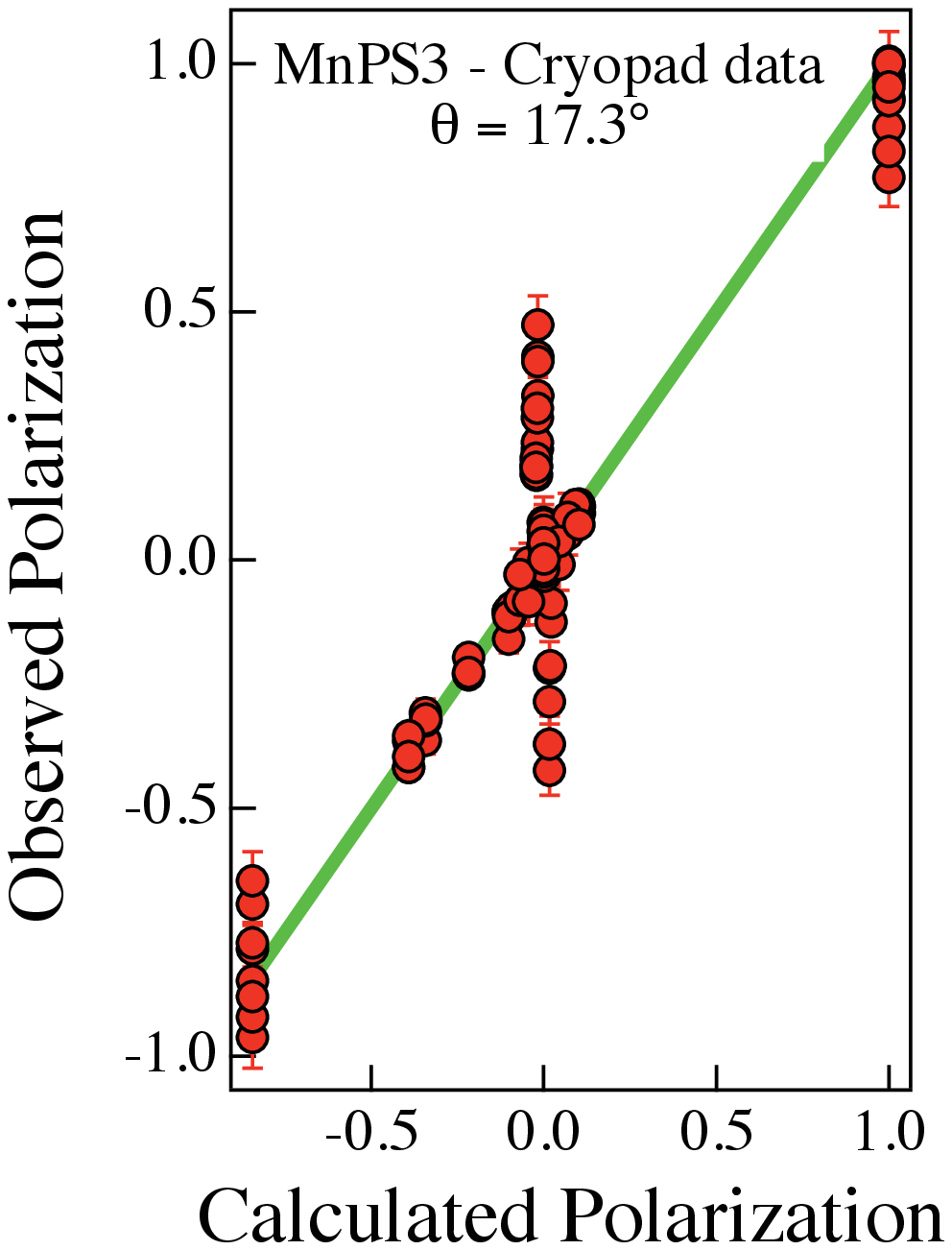}
\includegraphics[width=3.6cm]{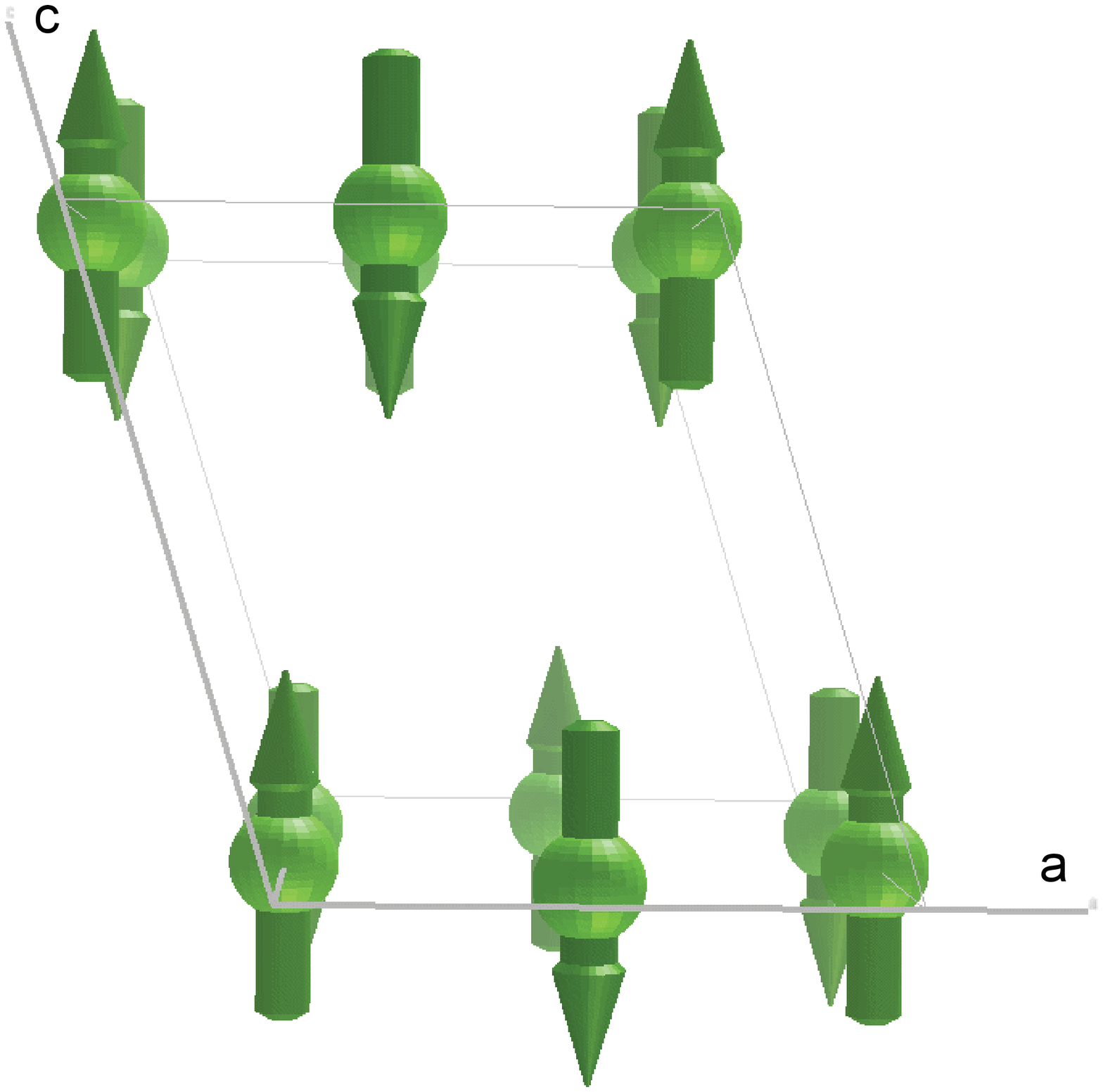}
\end{minipage}
\begin{minipage}[h]{.45\linewidth}
\includegraphics[width=4.0cm]{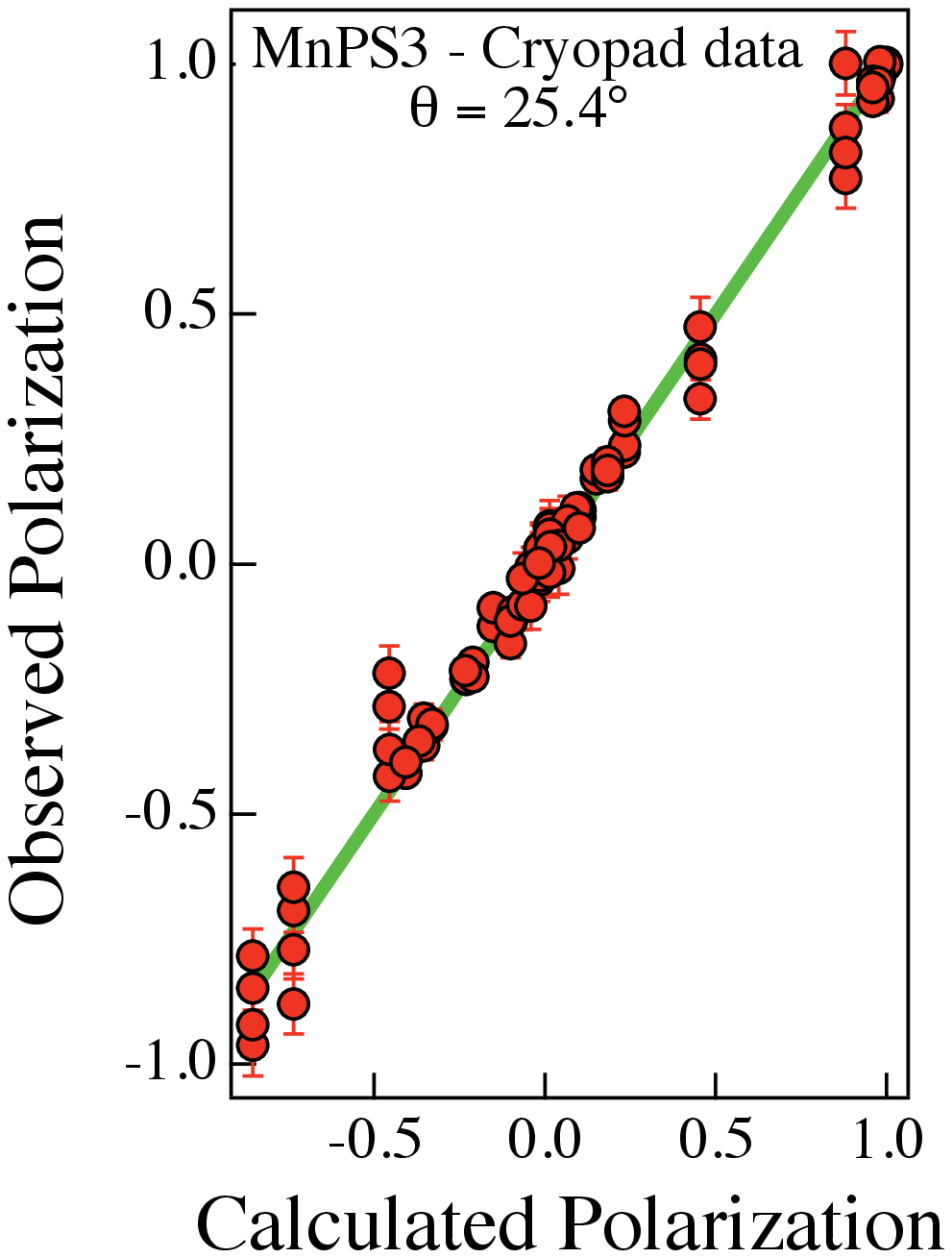}
\includegraphics[width=3.6cm]{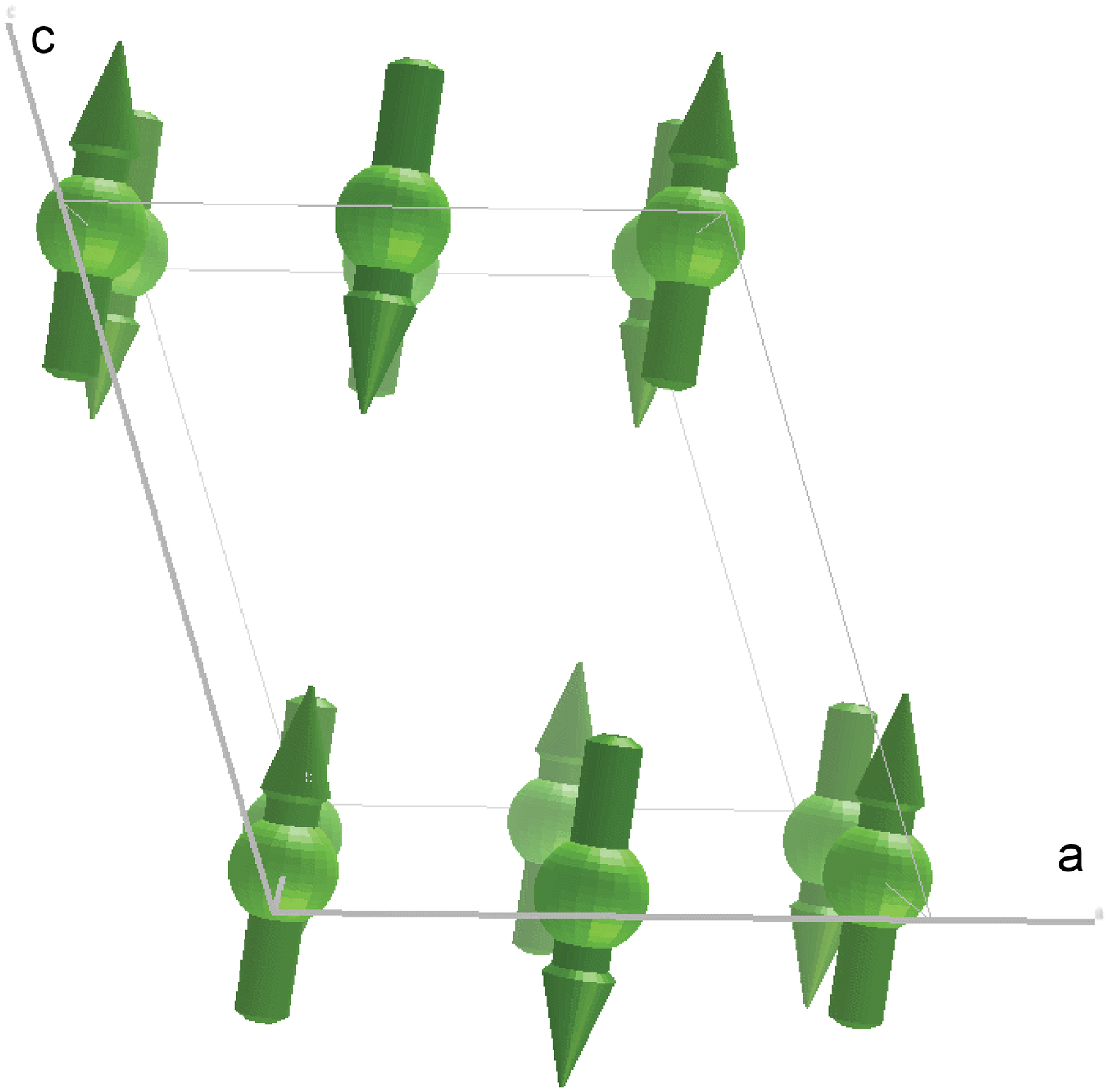}
\end{minipage}
\caption{(color online) Observed versus calculated polarization matrix elements for the published magnetic structure (left) where the magnetic moments are along $c^*$, and the revisited one (right) where the Mn$^{2+}$ moments lie at a finite angle of $\sim -8^{\circ}$ from this $c^*$ axis. The corresponding magnetic arrangements are shown in the bottom panels.}
\label{fits}
\end{figure}

\begin{figure}[h]
\begin{center}
\includegraphics[width=5.5cm]{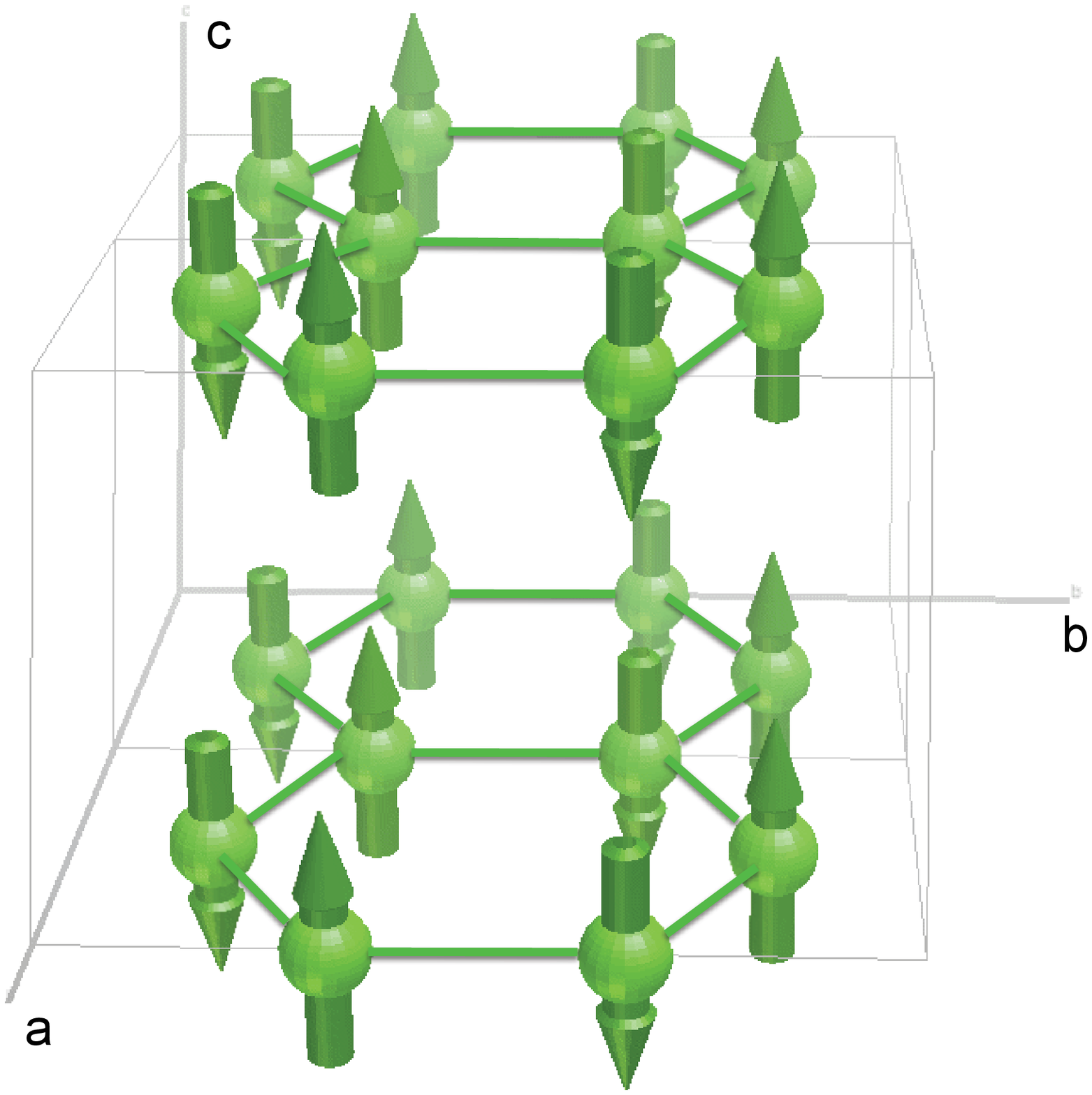}
\end{center}
\caption{(color online) Antiferromagnetic structure on the honeycomb lattice of Mn$^{2+}$.}
\label{magn}
\end{figure}

This magnetic structure description was further submitted to the stringent test of CRYOPAD.  The polarization matrices measured for a set of (0,$k$,$l$) reflections at 2 K without any ME annealing were refined. For each reflection, the 9 $P_{\ij}$ and the 9 $P_{-ij}$ terms (obtained by flipping the incident polarization)  were recorded during the experiment, thus eliminating possible systematic errors. The resulting 99 observations were fitted using the magnetic structure described before. The domain populations and the Mn$^{2+}$ magnetic moment amplitude were the only parameters of the fit. The result is shown in the left column of  Fig. \ref{fits}. In a second step, the angle $\theta$ between the moment direction and the $c$ axis in the $(c,a^*)$ plane was let free to vary. After refinement, $\theta$ was found equal to -25.4(2)$^{\circ}$:  taking into account the $\beta$ monoclinic angle, the moments form an angle of $\sim -8^{\circ}$ with respect to $c^*$ instead of being parallel to it. The result is shown in the right column of  Fig. \ref{fits}. This small change drastically improves the fit. The resulting magnetic structure is shown in Fig. \ref{magn}. The magnetic moment amplitudes are 4.43(3) $\mu_B$, slightly smaller than the expected saturated value of 5 $\mu_B$, but in good agreement with previous neutron works \cite{kurosawa}. In such a structure, there are only two 180$^{\circ}$  antiferromagnetic domains. Their volume fraction are $v_1 =55.3(3)\%$ and $ v_2 =44.7(3)\%$, i.e. close to an equirepartition. Usually, the domain repartition is quantified through the normalized volume difference $\eta={v_1-v_2\over v_1+v_2}=0.106(4)$ \cite{brown1}. The existence of a tilt of the magnetic moments with respect to natural symmetric directions (crystal axes) indicates the presence of competing anisotropies, result also suggested by a gap observed in the spin waves dispersion curves \cite{wildes1}. It could be due to dipolar and single-ion anisotropies, as found in other systems with Mn$^{2+}$ ions \cite{lhotel}.

The MnPS$_3$ magnetic point group is thus 2'/m where the inversion center is combined with time reversal. It is of particular interest because it allows a linear ME effect \cite{Tables}. Moreover, the ME tensor is constrained by symmetry to be non diagonal with non-equal off-diagonal elements. In the ($a$, $b$, $c$) frame, this tensor is:
\begin{equation*}
\begin{pmatrix}0&\alpha^{ME}_{12}&0\\ \alpha^{ME}_{21}&0&\alpha^{ME}_{23}\\ 0&\alpha^{ME}_{32}&0\end{pmatrix}
\end{equation*}

Given the possibility of a linear ME effect, we tried to characterize it, through its influence on the domain populations, by neutron spherical polarimetry. Indeed, in such a magnetic structure, the nuclear and magnetic structure factors are in phase quadrature (real structural part, imaginary magnetic part). A spherical polarization analysis using CRYOPAD is able to give information on the respective proportion of the two antiferromagnetic domains. The complete formalism is described in ref. \cite{brown1}. Calculations  show that the matrix elements $P_{XZ}$, $P_{ZX}$, $P_{XY}$, and $P_{YX}$ are highly sensitive to the domain population parameter $\eta$. 

\begin{table}[h]
\begin{center}
\begin{tabular}{ccccc}
\hline
$E$ (kV/m)   &  $\mu_{0}H$ (T) & $EH$ (Jm$^{-2}$ps$^{-1}$) & $P_{XZ}$      & $\eta$ \\
\hline
0   & 0 & 0 & -0.106(5) & 0.110(5)      	\\ 
271 $\hat c^*$   & 0 & 0 & -0.061(5) & 0.063(5)      	\\ 
\hline
271 $\hat c^*$   & 1.02 $\hat c^*$ & 0.221 & -0.877(4) & 0.913(5)      	\\ 
271 $\hat c^*$   & -1.02 $\hat c^*$ &-0.221 & 0.817(4) & -0.850(5)       	\\
214 $\hat c^*$   &  0.70 $\hat c^*$ & 0.120 & -0.274(5) & 0.285(5)       \\
85.7 $\hat c^*$   & 0.35 $\hat c^*$ & 0.024 & -0.119(5) & 0.124(5)      	\\  
\hline
271 $\hat c^*$   & 1.02 $\hat b^*$ & 0.221 & 0.913(4) & -0.950(5)      	\\ 
85.7 $\hat c^*$   & 0.35 $\hat b^*$ & 0.024 & 0.863(4) & -0.897(5)     	\\ 
42.9 $\hat c^*$   & 0.18 $\hat b^*$ & 0.006 & 0.613(5) & -0.637(5)       	\\  
21.4 $\hat c^*$   & 0.35 $\hat b^*$ & 0.006 & 0.339(5) & -0.353(5)       \\
21.4 $\hat c^*$   & 0.18 $\hat b^*$ & 0.003 & 0.484(5) & -0.503(5)       \\
10.7 $\hat c^*$   & 0.18 $\hat b^*$ & 0.002 & 0.384(5) & -0.399(5)       \\
5.3 $\hat c^*$   & 0.18 $\hat b^*$ & 0.001 & 0.500(5) & -0.520(5)       \\
5.7 $\hat c^*$   &- 0.18 $\hat b^*$ &- 0.001 & -0.451(5) & 0.469(5)       \\
\hline
\end{tabular}
\end{center}
\caption{Relative domain population $\eta$, after ME annealing for different orientations and amplitudes of the electric and magnetic fields. The first two lines corresponds to no ME annealing and annealing in $E$ only. The orientations of the fields are given by the unit vectors  $\hat c^*$ and $\hat b^*$ parallel to the $c^*$ and $b^*$ axes.}
\label{resdomain}
\end{table}

\begin{figure}[h]
\begin{center}
\includegraphics[width=7.5cm]{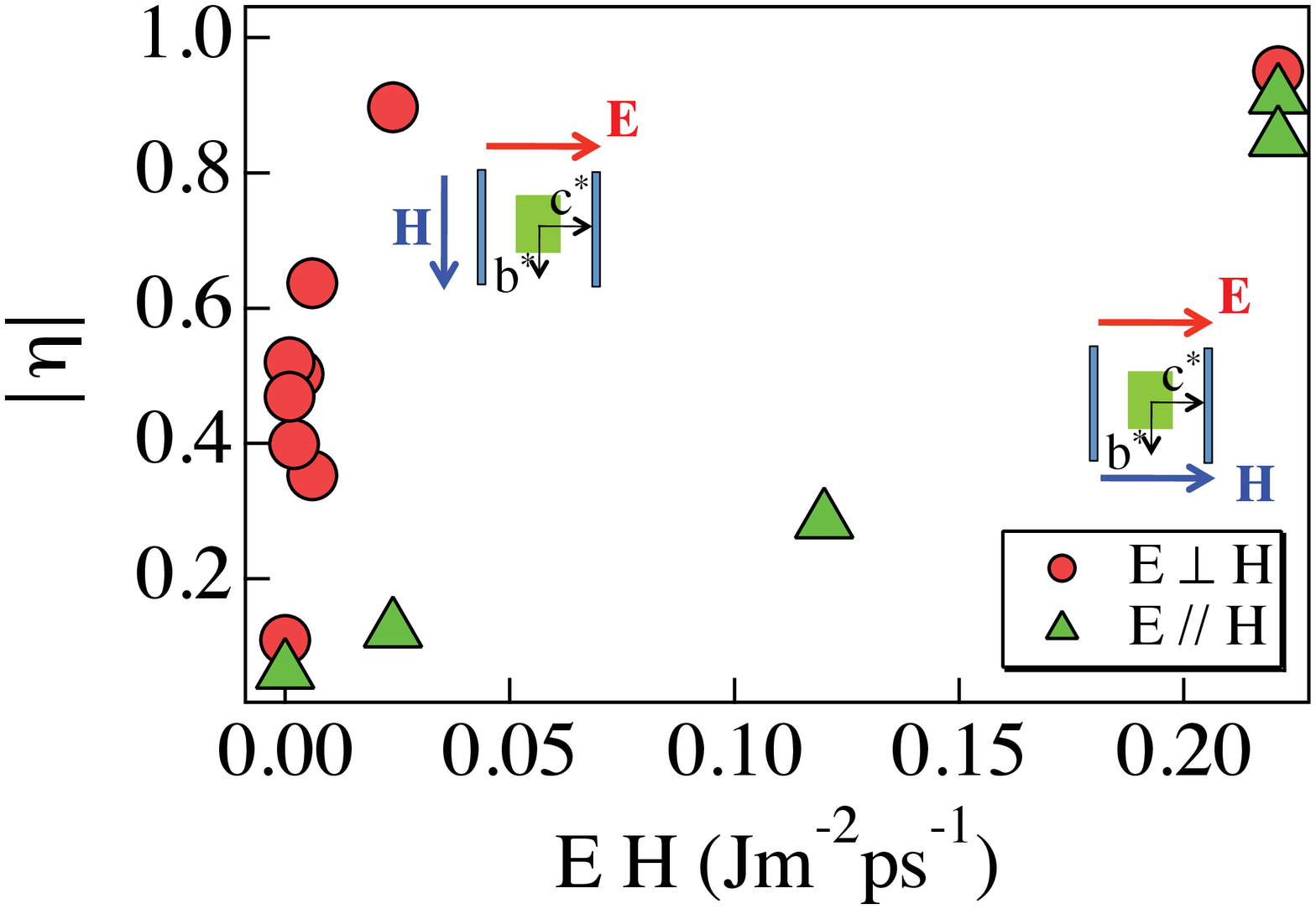}
\includegraphics[width=7.0cm]{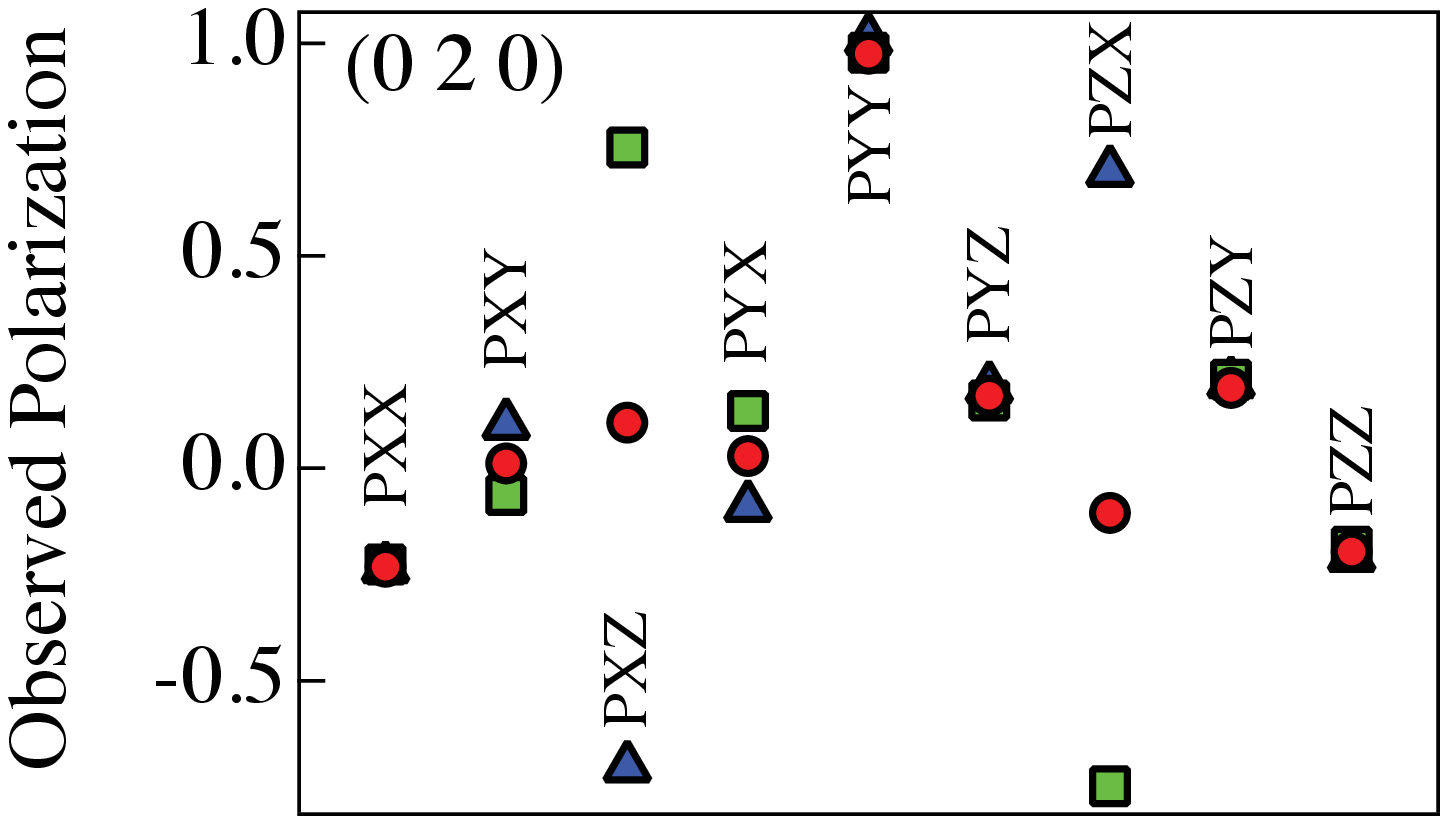}
\end{center}
\caption{Upper: Absolute value of the relative domain populations $\eta$ as a function of the product of the electric and magnetic fields for different respective orientations. Lower: polarization matrix elements of the (0 2 0) reflection for equi-populated antiferromagnetic domains (red dots) and for almost single domain 1 (blue triangles) or single domain 2 (green squares).}
\label{D3&polar}
\end{figure}

In this experiment, the MnPS$_3$ crystal has been submitted to electric and  magnetic fields. The electric field, with voltages up to 2 kV, was obtained via two Al electrodes separated from 7 mm surrounding the sample within the cryostat. It could only be applied along $c^*$ due to the lamellar sample shape. The tail of the cryostat could be placed within the gap of an electromagnet delivering magnetic fields as high as 1 T, parallel or perpendicular to the $E$ field direction {\it i.e.} along $c^*$ or $b^*$ respectively (see Fig. \ref{D3&polar}). For these two relative orientations of $E$ and $H$, the following combinations of ME tensor elements have been probed:  $$\alpha^{ME}_\Vert=\tan^2(\phi)\alpha^{ME}_{11}+({\tan(\phi)\over \cos(\phi)})(\alpha^{ME}_{13}+\alpha^{ME}_{31})+({1\over \cos(\phi)})^2\alpha^{ME}_{33}$$ $$\alpha^{ME}_\perp=\tan(\phi)\alpha^{ME}_{12}+{1\over \cos(\phi)}\alpha^{ME}_{32}$$ 
with $\phi=\beta-{\pi\over 2}=17.3^{\circ}$. Several ME annealings were produced by applying different electric and magnetic fields on the sample during cooling from the paramagnetic state (90 K) to the ordered phase (50 K). The fields were then switched off and the cryostat was positioned within CRYOPAD for the neutron polarization measurements. This procedure was repeated for each ($E$, $H$) set. The first measurements performed on several (0,$k$,$l$) reflections without ME annealing allowed to calculate precisely the domain population and to fix the different parameters. For subsequent measurements, only the $P_{XZ}$ matrix element of the (0,2,0) reflection was measured as the most sensitive term to domain populations (see Fig. \ref{D3&polar}). The domain populations were determined for increasing values of the product $EH$, for the parallel or perpendicular relative orientations of $E$ and $H$.

 It turned out that the domain distribution could be modified from $\sim$50\% of each domain without ME annealing to an almost single domain sample under the simultaneous action of $E$ and $H$ (results summarized in table \ref{resdomain} and Fig. \ref{D3&polar}). In the perpendicular geometry, the sample becomes rapidly single domain for fields product equal to $\sim$0.025 Jm$^{-2}$ps$^{-1}$ (typically 0.35 T and 86 kV/m). A single domain was also achieved for the parallel geometry but for field products ten times larger (typically 1 T and 270 kV/m). According to the symmetry of the ME tensor, only $\alpha^{ME}_\perp$ should be non-zero, and therefore acting on the domain population. The observed imbalance of the domain populations in the parallel geometry can actually be explained by a misalignment of both fields of the order of 5$^{\circ}$. The apparent dispersion of points for $E \perp H$ in  Fig. \ref{D3&polar} is of the same origin. Note that the reversal of either $E$ or $H$ favors a selection of the other antiferromagnetic domain. Finally, it has been checked that no ME annealing was obtained with an electric field alone. This experiment thus reveals the  particular form  of the linear ME tensor of MnPS$_3$, in agreement with the symmetry of the magnetic space group. 
 
 It is worth noting that this non-diagonal ME tensor allows in principle ferrotoroidicity, {\it i.e.} the alignment of toroidic moments. In the experiment geometry, it was impossible to investigate the antisymmetric part of the ME tensor proportional to the toroidic moment $T_k\propto (\alpha^{ME}_{ij}-\alpha^{ME}_{ji})\epsilon_{ijk}$ with $\epsilon_{ijk}$ the Levi-Civita symbol, as it was directly probed in GaFeO$_3$ \cite{popov}. However, a significant volumic toroidic moment could be calculated in the monoclinic unit cell: $T$=(0.0175, 0, -0.0027) $\mu_B$/\AA$^2$ \cite{ederer}. MnPS$_3$ thus provides with the first example of pure ferrotoroidicity (not associated with ferromagnetism and/or ferroelectricity) \cite{schmid2} with intrinsic parity-time reversal symmetry breaking. Ferrotoroidic domains should be present in MnPS$_3$ coinciding with the antiferromagnetic ones and manipulated conjointly \cite{schmid}. 

To summarize, a non-diagonal linear magnetoelectric effect has been established in the quasi 2D antiferromagnet manganese phosphorus trisulfide using polarized neutrons with spherical polarization analysis. The associated behavior is in agreement with the symmetries of the magnetic point group and also compatible with ferrotoroidicity. The ME domains population can be easily manipulated under reasonably weak fields through ME annealing. MnPS$_3$, with its simple magnetic structure and rather high ordering transition temperature, offers then a new model system to better understand magnetoelectricity and ferrotoroidicity.  

\acknowledgments We warmly thank A. Cano, E. Katz and P.J.~Brown for fruitful discussions, the latter also for the help with data analysis, and S. Vial for the technical support.

\end{document}